\documentclass[aps,prd,twocolumn,superscriptaddress,longbibliography,nofootinbib,preprintnumbers,amsmath,amssymb,floatfix,10pt]{revtex4-1}

\usepackage{epsfig}
\usepackage{amsmath,amssymb,amsfonts}
\usepackage{color}
\usepackage{hyperref}
\usepackage{mathrsfs}
\usepackage{slashed}
\usepackage{graphics}

\newcommand{\be}{\begin{equation}}
\newcommand{\ee}{\end{equation}}
\newcommand{\bi}{\begin{itemize}}
\newcommand{\ei}{\end{itemize}}
\newcommand{\bea}{\begin{eqnarray}}
\newcommand{\eea}{\end{eqnarray}}

\newcommand{\bra}[1]{\langle\,#1\,|}          
\newcommand{\ket}[1]{|\,#1\,\rangle}          

\newcommand{\ud}{\mathrm{d}}

\newcommand{\LCm}{{\scriptscriptstyle -}} 
\newcommand{\LCp}{{\scriptscriptstyle +}}
\newcommand{\LCpm}{{\scriptscriptstyle \pm}}
\newcommand{\LCmp}{{\scriptscriptstyle \mp}}

\newcommand{\LCperp}{{\scriptscriptstyle \perp}}

\definecolor{gold}{rgb}{1,0.75,0}

\begin{document}

\title{Strong field effects in laser pulses: the Wigner formalism.}

\author{Florian Hebenstreit}
\email[]{florian.hebenstreit@uni-graz.at}
\affiliation{Institut f\"ur Physik, Karl-Franzens Universit\"at Graz, A-8010 Graz, Austria}

\author{Anton Ilderton}
\email[]{anton.ilderton@physics.umu.se}
\affiliation{Department of Physics, Ume\aa\ University, SE-901 87 Ume\aa, Sweden}

\author{Mattias Marklund}
\email[]{mattias.marklund@physics.umu.se}
\affiliation{Department of Physics, Ume\aa\ University, SE-901 87 Ume\aa, Sweden}
\author{Jens Zamanian}
\email[]{jens.zamanian@physics.umu.se}
\affiliation{Department of Physics, Ume\aa\ University, SE-901 87 Ume\aa, Sweden}
%
\begin{abstract}
We investigate strong field vacuum effects using a phase space approach based on the Wigner formalism. We calculate the Wigner function in a strong null-field background exactly, using lightfront field theory. The Wigner function exhibits the distinct features of strong field QED: in particular we identify the effective mass in a laser pulse, compare it to the well-known mass shift in a periodic plane wave and identify signals of multi-photon absorption and emission. Finally, we show how to extend our results to describe vacuum pair production in colliding laser pulses. 
\end{abstract}
\maketitle
\section{Introduction.}
Vacuum phenomena generated by strong external fields are currently a topic of considerable interest, due to the prospects provided by infrastructure projects such as the European XFEL \cite{Ringwald:2001ib} and the ELI facility \cite{Dunne:2008kc, Heinzl:2008an}. The latter is expected to produce laser pulses with an intensity of $10^{24}$ W/cm$^2$ and peak electric field strengths of $10^{14}$ V/cm. These fields can produce strongly relativistic ionisation effects, polarise the vacuum and so generate nontrivial vacuum refractive indices \cite{Marklund:2006my, Heinzl:2006xc,Heinzl:2006pn}, be used to search for new light particles \cite{Gies:2008wv}, and, perhaps, pull electron-positron pairs out of the vacuum \cite{Bulanov:2010ei}.

Such strong field physics has been investigated using scattering calculations \cite{Harvey:2009ry}, exact solutions \cite{Dunne:1998ni}, semiclassical techniques \cite{Brezin:1970xf, Kim:2000un, Piazza:2004sv, Dumlu:2010ua}, Monte Carlo simulations \cite{Gies:2005bz} and quantum kinetic equations \cite{Alkofer:2001ik, Blaschke:2005hs, Hebenstreit:2009km}. A framework which includes quantum kinetic theory is provided by the Wigner function, the quantum analogue of a probability distribution. It is a well-established tool in quantum optics \cite{Lee:1995}, and has  been applied to transport phenomena in plasmas and condensed matter systems \cite{Elze:1986qd, Haug}, and to pair production in both strong electric and colour fields \cite{BialynickiBirula:1991tx, Hebenstreit:2010vz, Levai:2009mn}. (For connections with other approaches to pair production, see \cite{Dumlu:2009rr, Fedotov:2010ue}.) The Wigner formalism has the advantage of offering a phase-space description of quantum systems with direct access to real time processes. For a thorough introduction see \cite{Zhuang:1995pd}.

One of the challenges of this approach lies in its interpretation: quantum effects can generate negative values of the Wigner function which have no probabilistic interpretation.  Nevertheless, the Wigner function is an observable, and indeed has been measured in various (non-relativistic) optical and atomic setups \cite{Kurtsiefer:1997,Breitenbach:1997,Bertet:2002}. Consequently, it is of great importance to obtain a thorough analytical understanding of the Wigner function. In this paper we therefore calculate the exact (relativistic) electron-positron Wigner function, in the presence of a strong background field, and give a complete description of its properties. Our chosen backgrounds, modeling high intensity lasers, are null-fields, i.e.\ plane waves.  This automatically includes effects due to strong magnetic, as well as electric, fields. Although we neglect transverse size effects, we take explicit account of effects due to finite pulse duration by considering finite {\it longitudinal} extent of the fields.

The most well known effect of a plane wave on an electron is the generation of an intensity dependent effective mass \cite{Sengupta:1952}. This is only well understood for very long laser pulses modelled by periodic plane waves (i.e.\ those of infinite duration), where rapid changes in the particle's energy and momentum, due to oscillations of the fields, average out to a mass shift. Modern laser pulses are, however, only a few cycles long. In order to examine the impact of finite beam geometry, we will compare results for periodic plane waves with those for more realistic pulse profiles. Using the Wigner function, we will identify the effective mass for an {\it arbitrary} profile.

This paper is organised as follows: we begin in Sect.~\ref{intro1} by briefly reviewing the gauge invariant Wigner function in both covariant and equal time formulations. We then recap the behaviour of charges in a plane wave, in Sect.~\ref{intro2}. In Sect.~\ref{WVsect}, we construct the Wigner function in a plane wave background exactly, and identify the effective mass in a pulse. We examine in detail the Wigner function for both oscillatory and pulsed fields, identifying effects due to the effective mass, multi-photon contributions and quantum corrections. Finally, we add a second laser pulse and describe how our results may be extended to include pair production.
\section{The Wigner function.}\label{intro1}
The electron-positron Wigner function in an external electromagnetic field $F_{\mu\nu}$ is constructed from the field density in the vacuum state $\ket{0}$:
\be\label{U4}
  \bra{0} [\psi(x+\tfrac{y}{2}),\psi^\dagger(x-\tfrac{y}{2})]\ket{0} =: \mathcal{U}_4(x+\tfrac{y}{2},x-\tfrac{y}{2}) \; ,
\ee
where $x$  and $y$ are the `centre of mass' and `relative' coordinates, respectively. The covariant Wigner function itself, $\mathcal{W}_4$, is obtained from $\mathcal{U}_4$ by Fourier transforming with respect to the relative co-ordinate, trading $y^\mu$ for a momentum $p_\mu$ whose physical interpretation will be given directly:
\be
  \label{Wig}
  \mathcal{W}_4(x,p)=\int\! \ud^4y\ e^{ip.y/\hbar}\, e^{ie\int\! \ud z.A(z)/\hbar}\, \mathcal{U}_4 \; .
\ee
The Wilson line in (\ref{Wig}) makes $\mathcal{W}_4$ {\it gauge invariant} and is evaluated along the straight line path from $x-y/2$ to $x+y/2$ \cite{Elze:1986qd}. Expanding in powers of $\hbar$, one can show that the classical part of $\mathcal{W}_4$ is always constrained to obey $p^2-m^2=0$, and hence $p_\mu$ may be identified with the kinematic momentum. As we will calculate {\it exactly} in $\hbar$, we set it equal to one from here on. The Wigner function obeys a manifestly gauge invariant version of the Dirac equation \cite{Zhuang:1995pd}, which we will return to in Sect.~\ref{PPsect}. Consequently, there are at least two approaches to constructing Wigner functions. First, by solving this differential equation, which is a manifestly gauge invariant approach. Alternatively, one can fix the gauge, quantise the theory and construct the expectation value in (\ref{U4}) directly. We give here a simple example of the latter method: the free Wigner function is easily found by quantising the free fermion theory and calculating the commutator in the free vacuum state. One obtains
\be
  \label{W4Free}
  \begin{split}
	\mathcal{W}_4(x,p)_\text{free} & = 2\pi\,\delta(p^2-m^2) [\slashed{p}+m] \\
	&\equiv W_4(x,p)_\text{free}[\slashed{p}+m]\; ,
  \end{split}
\ee
so that the free Wigner function is spatially homogeneous and tells us that free particles are on-shell. The final equality {indicates} a separation of $\mathcal{W}_4$ into spin and scalar contributions -- in fact, $W_4$ is the Wigner function for {\it scalar} particles, as is easily found by calculating the free scalar density
\be\label{U4}
  U_4(x+\tfrac{y}{2},x-\tfrac{y}{2}) := \bra{0} \{\phi(x+\tfrac{y}{2}),\phi^\dagger(x-\tfrac{y}{2})\}\ket{0} \; ,
\ee
and performing the analogue of the transformation (\ref{Wig}).

One often also considers the equal time Wigner function $\mathcal{W}_3$. This is defined as the energy average of $\mathcal{W}_4$, i.e.\ it is obtained by integrating out $p_0$. For the free theory one therefore has:
\be
  \label{W3Free}
  \mathcal{W}_3(t; {\bf x},{\bf p})_\text{free} := \int\! \frac{\ud p_0}{2\pi}\ \mathcal{W}_4(x,p)_\text{free}=\frac{   {{\bf {p}}}\!\!\!/+m              }{\sqrt{{\bf p}^2+m^2}} \; .
\ee
At this point we introduce lightfront co-ordinates $x^\LCpm=x^0\pm x^3$, $x^\LCperp = \{x^1,x^2\}$ which will be  useful below. In lightfront field theory (for reviews see \cite{Brodsky:1997de, Heinzl:2000ht} and references therein) one quantises at equal lightfront time $x^\LCp$. One simplification of this approach is that only half of a Dirac spinor's components represent dynamical degrees of freedom \cite{Brodsky:1997de, Heinzl:2000ht}. Introducing the orthogonal projection operators $\Lambda^\LCpm = \tfrac{1}{4}\gamma^\LCmp \gamma^\LCpm$, the dynamical fermion field is $\psi_{\scriptscriptstyle (+)}\equiv\Lambda^\LCp\psi$. On the other hand, $\psi_{\scriptscriptstyle (-)}\equiv\Lambda^\LCm\psi$ is a constrained field which is determined by $\psi_{\scriptscriptstyle (+)}$. In this approach we therefore construct the Wigner function from the commutator of the dynamical field $\psi_{\scriptscriptstyle (+)}$ in the lightfront vacuum. The density becomes
\be
  \mathcal{U}_4 = \bra{0} \big[ \psi_{\scriptscriptstyle (+)}(x + \tfrac{y}{2}) , \psi_{\scriptscriptstyle (+)}^{\dagger} (x - \tfrac{y}{2}) \big] \ket{0} \; 
\ee
and the Wigner function is again found via (\ref{Wig}). The spin structure in this picture leads to a very simple relation between the spinor and scalar Wigner functions: one finds that the free Wigner function is given by the scalar result multiplied by the physical projector, so	
\be\label{relation}
	\mathcal{W}_4(x,p)_\text{free} = 2 p_\LCm \Lambda^\LCp W_4(x,p)_\text{free} \;,
\ee
where, of course, the free scalar Wigner function matches that found in the usual approach, see (\ref{W4Free}). As we will see, relation (\ref{relation}) also extends to the interacting theory. For future use we note that, using lightfront co-ordinates, the free scalar Wigner function may be written
\be
  \label{W4Freelight}
	W_4(x,p)_\text{free} = \frac{\pi}{2| p_\LCm |}\delta\bigg(p_\LCp - \frac{{p}_\LCperp^2+m^2}{4p_\LCm}\bigg) \;,
\ee
which simply exhibits the mass-shell condition in lightfront co-ordinates. It follows that the equal {\it lightfront time} Wigner function in the free fermion theory is (writing $\mathsf x \equiv \{x^\LCm, x^\LCperp\}$, $\mathsf p \equiv \{p_\LCm, p_\LCperp\}$)
\be\label{W3plus}
	\mathcal{W}_3^\LCp(x^\LCp; {\mathsf x}, \mathsf{p})_\text{free} := \int\! \frac{\ud p_\LCp}{2\pi}\ \mathcal{W}_4(x,p)_\text{free} = 2 p_\LCm  \Lambda^\LCp \cdot \frac{1}{4|p_\LCm|}\;,
\ee
where the `$\cdot$' simply illustrates the separation between spin and scalar contributions. 

\section{Plane waves and the mass shift.}\label{intro2} 
We now review the effects of a plane wave background on particle motion and QED scattering events. The field strength of a general plane wave is represented by
\be
  F_{\mu\nu} = \dot{f}_j(k.x)  ( k_\mu a^j_\nu -a^j_\mu k_\nu ) \ ,
\ee
where $k^2=0$, $j$ is summed over the two transverse directions and a dot is differentiation with respect to $k.x$ (for convenience). The polarisation vectors obey $a^i.k=0$, $a^i.a^j=-{(}m^2a^2/e^2{)}\delta^{ij}$, which defines an invariant, dimensionless amplitude $a$. Since $k_\mu$ is a lightlike direction we take $k_\mu = \omega(1,0,0,1)$ so that $k.x =\omega x^\LCp$ is proportional to lightfront time.

Consider first the Lorentz equation for the classical orbit of a charge in this background.  This is exactly solvable: $k.p$ is conserved and hence the charge's proper time is proportional to lightfront time $x^\LCp$. In long pulses modelled by periodic fields, i.e.\ plane waves of infinite extent, the average, or `quasi' momentum of the charge over a laser cycle is of particular importance: it is on-shell, but for a shifted mass $m_*$, i.e. $\langle p \rangle^2=m_*^2$. For periodic, infinite plane waves -- IPWs from here on -- the mass shift is $m_*^2=m^2(1+a_0^2)$, where we have introduced the common measure of laser intensity $a_0^2$, which equals $a^2$ for circular polarisation and $a^2/2$ for linear polarisation \cite{Heinzl:2008rh}. Currently, $a_0\sim\mathcal{O}(10)$, while for ELI-strength lasers $a_0\simeq10^3-10^4$. 

The effective mass is only well-understood in an IPW, where the relevant average is taken over a laser cycle. The IPW model was assumed in the analysis of the SLAC E144 pair production experiment \cite{Bamber:1999zt}, in which the laser pulse contained around $10^3$ cycles. To model modern femtosecond duration pulses, it is necessary to look at pulse profiles with finite extent in $k.x$. Periodicity is then lost, and thus it is unclear what average momentum is important, nor if any mass-shift remains. We will address this shortly. Suppose for a moment that some average $\langle\ \rangle$ over lightfront time can be identified, be it the cycle average or otherwise. Then, from the classical orbit, one finds
\be \label{var}
  \langle p \rangle^2 = m^2(1 + a^2\mathsf{var}^2[f_j]) \;,
\ee
where $\mathsf{var}^2[f]=\langle f^2\rangle - \langle f\rangle^2$. Thus, the mass shift in periodic fields is given by the {\it variance of the integrated field strength over a laser cycle}. It does not seem that this statement, which will be useful below, is well known, although we believe the first appearance of the variance expression is in \cite{Brown:1964zz}.

\begin{figure}
\centering
\includegraphics[width=0.75\columnwidth]{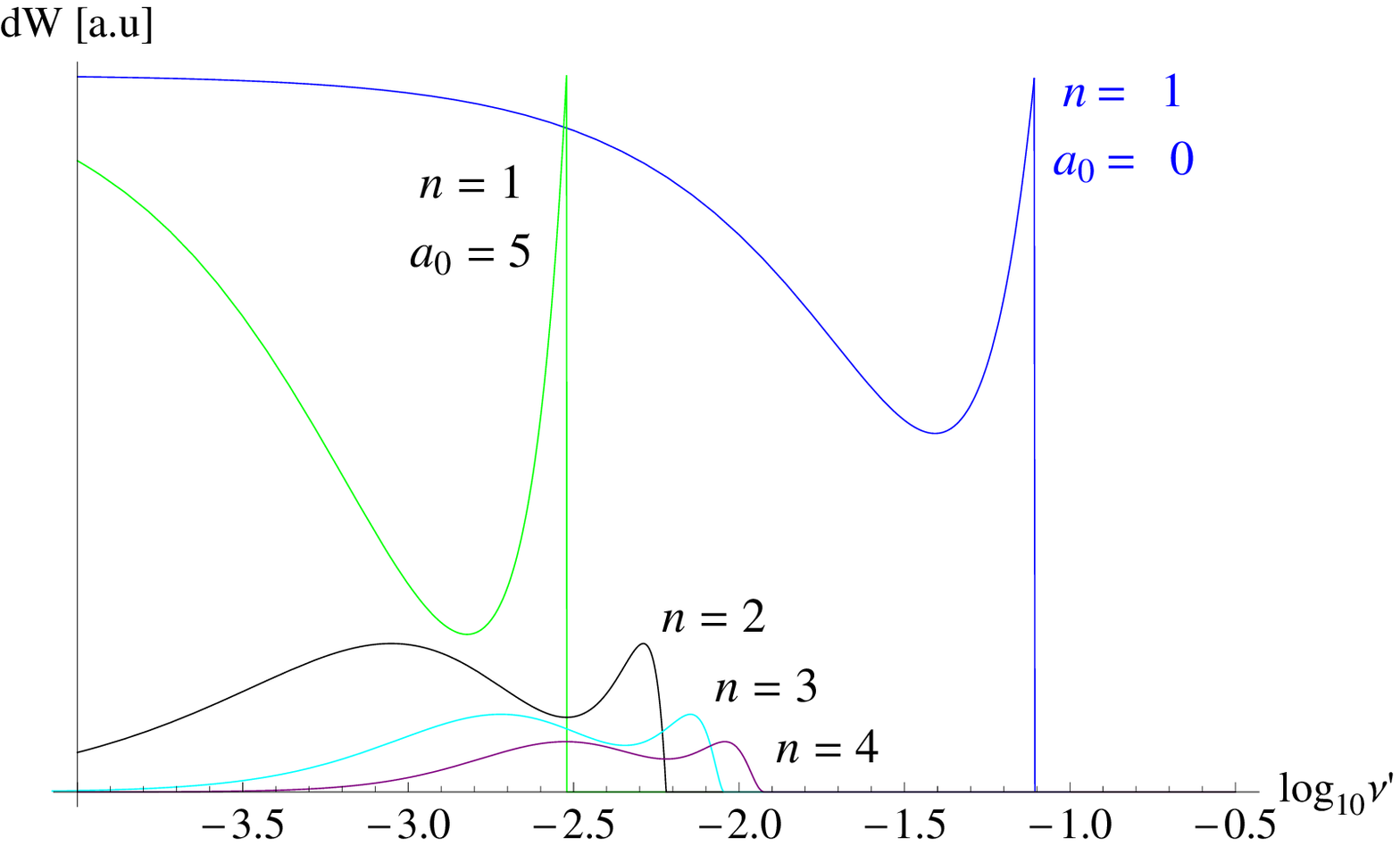} 
\includegraphics[width=0.75\columnwidth]{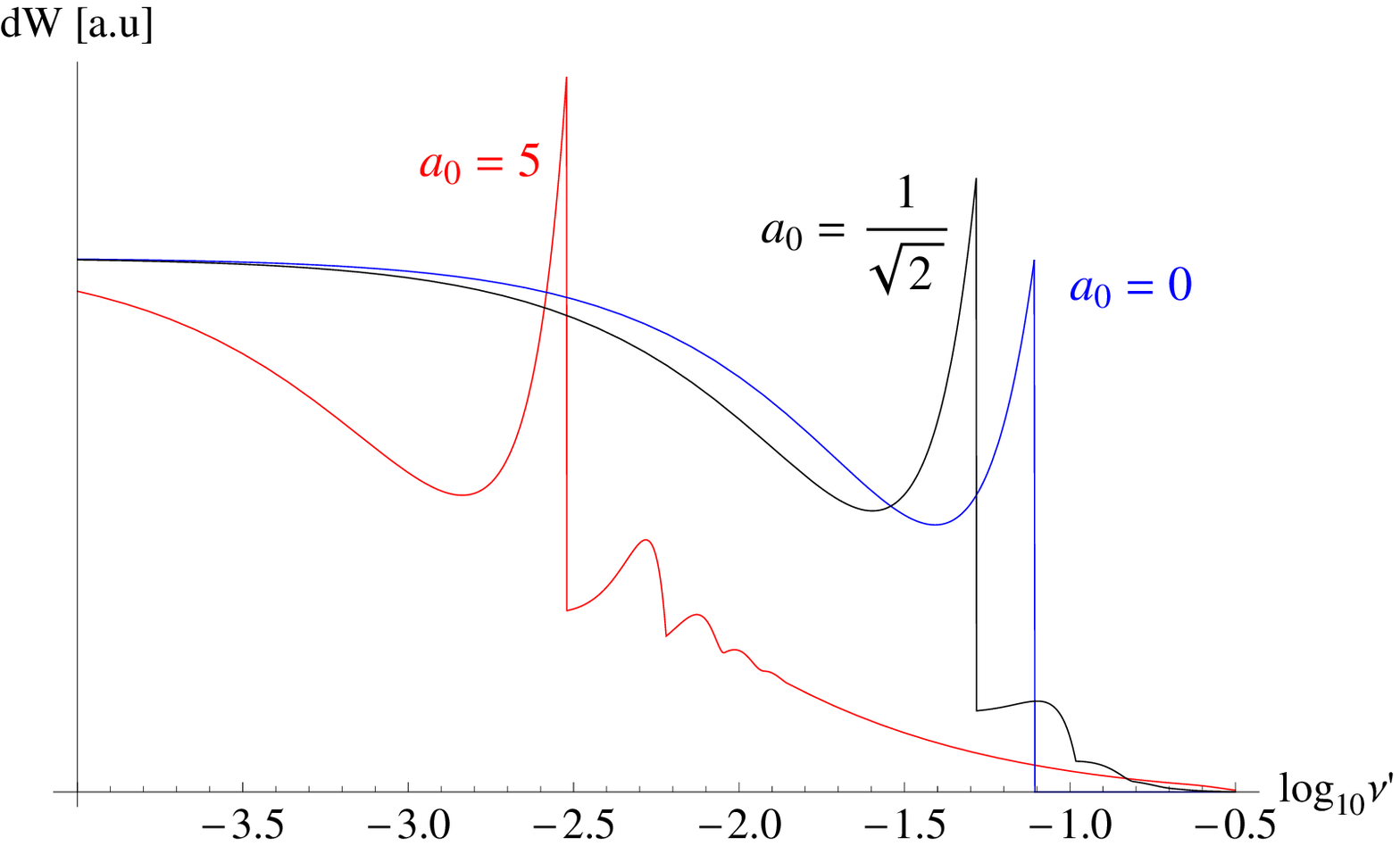}
\caption{\label{FIG:NLC} Emission spectrum for photons scattered from an electron-laser collision, as a function of scattered photon frequency $\nu'=\omega'/m$. {\it Top}: $n=1$ to $4$ laser photon contributions (harmonics) at an intensity $a_0^2=25$, compared to the ordinary Compton scattering spectrum ($n=1$ at $a_0=0$). {\it Bottom}: the full spectrum for $a_0^2=25$ and $a_0^2=1/2$, summed over all photon numbers.}
\end{figure}
Going to QED, one finds that scattering amplitudes have kinematic support on the conservation of quasi-momentum: hence, for example, electron-positron pair production stimulated by incoming probe photons is described in terms of the production of mass $m_*$ pairs \cite{Nikishov:1964zz,Nikishov:1964zza,Narozhnyi:1964}.  In addition to mass-shift effects, QED scattering processes in periodic plane waves exhibit effects due to multi-photon absorption from the background. In nonlinear Compton scattering, i.e.\ the emission of a photon from an electron in a laser field, the mass shift is responsible for a red-shift of the kinematic Compton edge, while multi-photon effects allow for higher harmonics, that is scattered photons with energy higher than the edge value \cite{Harvey:2009ry}. The full scattering process is a sum over all multi-photon contributions. This is illustrated in Fig.~\ref{FIG:NLC}. (See \cite{Heinzl:2009nd,Seipt:2010ya,Mackenroth:2010jr} for corrections due to finite size effects.)  This completes our discussion of the most well-known effects in periodic plane waves. We now turn to a closer investigation of both periodic and pulsed laser field effects, using the Wigner function.
\section{An exact Wigner function}\label{WVsect}
In this section we calculate the Wigner function in a plane wave field exactly. The Klein--Gordon and Dirac equations are exactly solvable in this background \cite{Volkov:1935}, at least when one works in lightfront gauge $k.A\sim A^\LCp=0$, when the gauge potential for our $F_{\mu\nu}$ becomes
\be
\label{A}
	A_\mu(k.x) = f_j(k.x)a^j_\mu \;.
\ee
In this setup, we have a correspondence with, e.g., \cite{Hebenstreit:2010vz}, which takes $F_{\mu\nu}$ to be a spatially homogeneous, time dependent electric field. Instead of ordinary time {$t$}, we have front time $x^\LCp$. Instead of temporal gauge, $A^0=0$, we have lightfront gauge $A^\LCp=0$. The $t$  dependent electric field is replaced by $x^\LCp$ dependent electric and magnetic fields, and all fields are homogeneous in their corresponding 3-spaces. The real difference is that the former field produces pairs, the latter does not. We will include a second plane wave, with which we can generate pairs, in a later section. 

\begin{figure}[t]
\centering
\includegraphics[width=0.76\columnwidth]{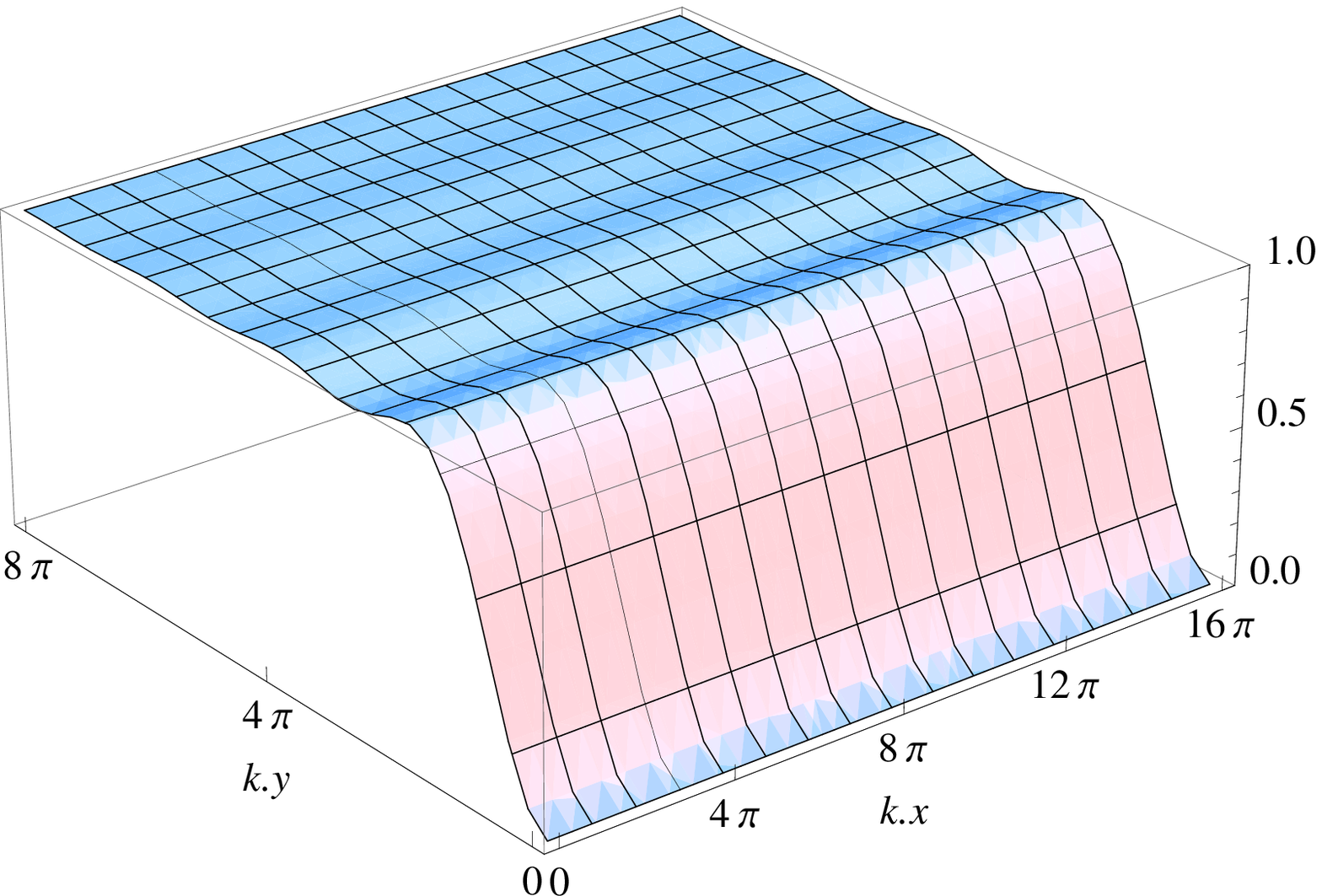}
\includegraphics[width=0.75\columnwidth]{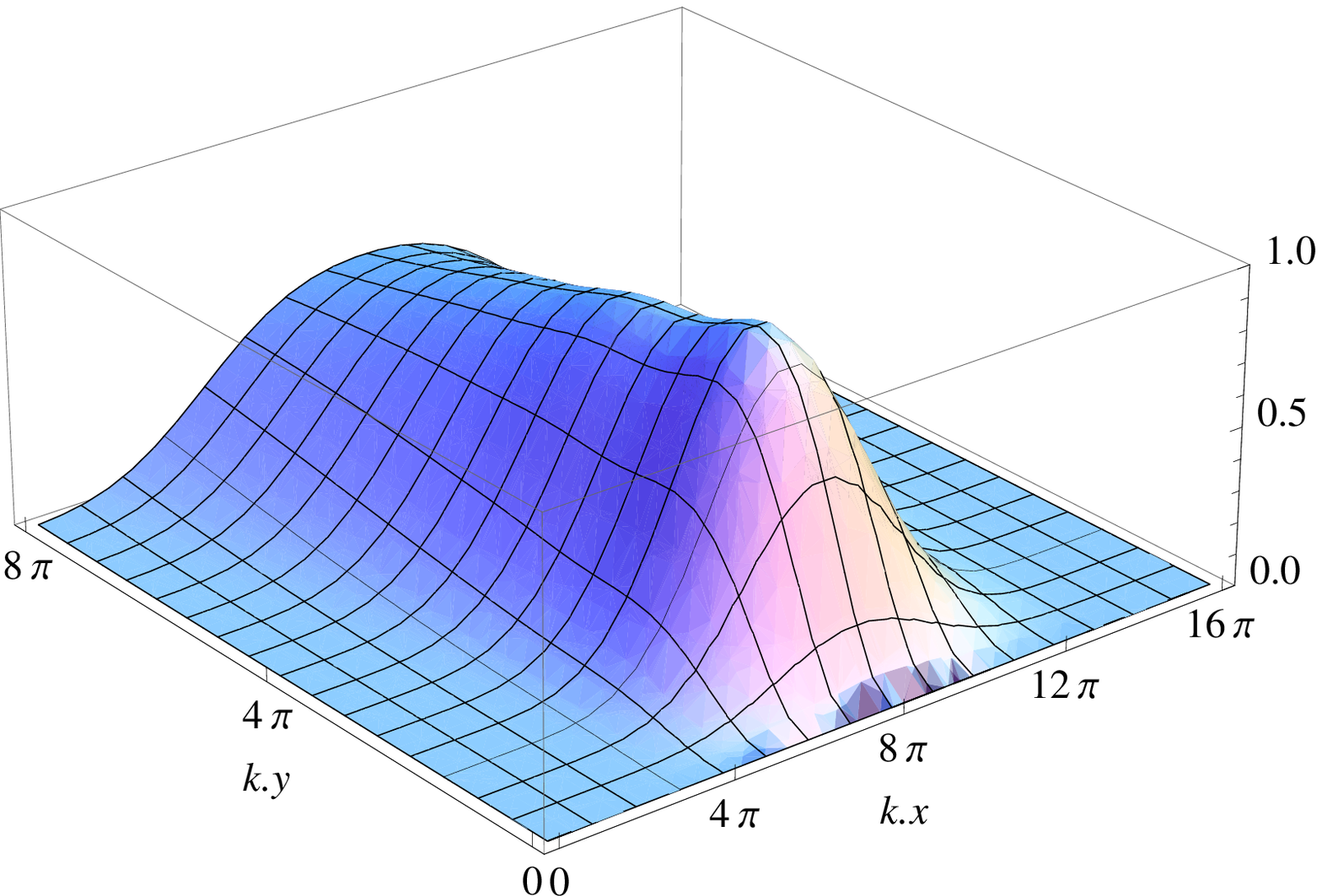}
\caption{\label{FIG:SHIFT} The mass shift $M^2/m^2-1$ in units of $a^2$, i.e.\ $\mathsf{var}^2[f_j]$. {\it Top}: in a circular polarised IPW the mass shift tends to {$1$}, i.e.\ $M^2\to m_*^2$, as $k.y$ increases. {\it Bottom}: in the 8-cycle pulse (\ref{PulseChoice}), $M$ reduces to the free mass $m$ for large $k.x$ or $k.y$.}
\end{figure}

To proceed, one quantises at equal lightfront time and calculates the field density in the lightfront vacuum state. Just as in the free case (\ref{relation}), one finds that the electron-positron Wigner function is given by the a scalar contribution multiplied by the background-independent spin contribution $2p_\LCm\Lambda^\LCp$. For this reason, we restrict the remaining discussion to the interesting scalar part (which is again the Wigner function for scalar fields). One finds that the scalar $U_4$ is
\be  \label{VolkU}
\begin{split}
	U_4 &= \int\!  \frac{\ud^4q}{{(2\pi)^3}} \ \delta(q^2-m^2) e^{-iq.y} \\
	&\times\exp\bigg[ \frac{-i}{2k.q}\int\limits_{k.(x-y/2)}^{k.(x+y/2)}\!\!\!\ud s\ [ 2e A(s).q - e^2 A^2(s) ] \bigg] \;,
\end{split}
\ee
where we recognise the usual Volkov phase factor \cite{Volkov:1935}. It is possible to perform the $y^\LCm$ and $y^\LCperp$ integrals in (\ref{Wig}) exactly, to obtain the compact, and exact, expression
\be \label{result}
	W_4(x,p) = \frac{1}{4|p_\LCm|} \int\!\ud y^\LCp \exp\bigg[ i y^\LCp\bigg(p_\LCp - \frac{p_\LCperp^2 + M^2}{4p_\LCm} \bigg)\bigg]  \;,
\ee
where $M^2(k.x,k.y) := m^2( 1+ a^2\mathsf{var}^2[f_j])$, with the average taken over the separation of the fields, i.e.\
\be
	  \langle f \rangle := \int\limits_{k.(x-y/2)}^{k.(x+y/2)}\! \!\! \!\ud s\ \frac{f(s)}{k.y} \,.
\ee
When the fields turn off, $M\to m$ and the $y^\LCp$ integration correctly recovers the delta function (\ref{W4Freelight}), imposing the free mass-shell condition. Hence (\ref{result}) describes a deformation of this condition in which the free mass is replaced with an {\it effective mass} $M$ of the IPW form (\ref{var}): hence, our Wigner function gives us the explicit form of the effective mass for an arbitrary pulse profile. This is corroborated by \cite{Kibble:1975vz}, where precisely this $M$ appears in the gauge invariant part of the electron propagator. (See \cite{Bonitz} for related non-relativistic results.) Let us compare the effective mass in a circularly polarised IPW ($ f_1=\sin{(k.x)}$, $ f_2=\cos{(k.x)}$) with that in an $n$-cycle pulse, where \cite{Mackenroth:2010jk}
\be\label{PulseChoice}
\begin{split}
	f_1(k.x) &= \sin^4(\tfrac{k.x}{2n})\sin(k.x)\;, \\
	f_2(k.x) &= \sin^4(\tfrac{k.x}{2n})\cos(k.x)\;,
\end{split}
\ee
for $0\leq k.x \leq 2\pi n$ and zero otherwise. A pulse with $n=8$, for example, corresponds to 32 fs lab duration at optical frequency. The effective mass in the IPW is 
\be  \label{Mcirc}
	M^2_\text{circ}/m^2=1 + a^2 - a^2\text{sinc}^2(\tfrac{k.y}{2})  \;,
\ee
which is independent of $k.x$ (linear polarisation introduces periodicity in $k.x$). We do not write down the lengthy expression for the pulsed $M$ explicitly. The effective mass in both the IPW and the pulse (\ref{PulseChoice}) are plotted in Fig.~\ref{FIG:SHIFT}. The IPW result is shown in the top panel, with the interesting behaviour in the $k.y$ direction: at $k.y=0$, the effective mass is the free mass. As $k.y$ approaches $2\pi$, i.e. a front time duration corresponding to one laser cycle, $M^2$ increases quickly toward $m^2(1+a^2)$, and converges toward this value as $k.y\to\infty$.  This is of course the IPW $m_*^2$ associated with the cycle average: hence, as we know, a particle seeing infinitely many laser cycles behaves as if its rest mass were $m_*$.

In the bottom panel, we see that the effective mass in a pulse behaves quite differently. For large $k.x$, $M\to m$, as one sits far outside the pulse. As the front time separation $k.y$ increases from zero, $M$ rises inside the pulse, but then falls back to zero for increasing $k.y$. This reflects the statement that any massive particle in the field will leave the pulse after a finite time (front or instant), so that eventually the effective mass reduces to the free mass. In particular, we see rather cleanly that scattering processes in a {\it pulse}, which are asymptotic in $k.y$, must be sensitive to the free mass.

\subsection{Monochromatic limit}
\begin{figure}[t!]
\includegraphics[width=\columnwidth]{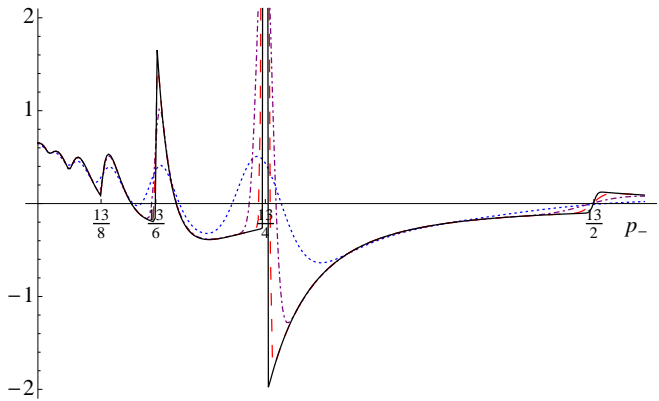} 
\caption{\label{FIG:FIRST} The Wigner function, [a.u.], plotted at $p_\LCperp=0$, $p_\LCp=2$, $a_0^2=25$ as a function of $p_\LCm$. We clearly observe convergence for decreasing values of $\epsilon$ ($=10^{-4}$ blue/dotted line, $10^{-5}$ purple/dot-dashed, and $10^{-6}$ red/dashed) towards the final result (solid). The peak sits at the shifted mass-shell condition, and the photon emission and absorption numbers discussed in the text are also marked on the axes.}
\end{figure}

We now perform the $y^\LCp$ integration in (\ref{result}) and turn to the Wigner function itself, beginning with the IPW result. To do so we add a Gaussian regulator $\exp[-\epsilon\, (k.y)^2]$ under the integral of (\ref{result}) and take $\epsilon\to0$. (We have confirmed that our results are unchanged using a different regulator.) Like the free solution, our $W_4$ is spatially homogeneous, as follows from (\ref{Mcirc}) being independent of $k.x$. $W_4$ is plotted in Fig.~\ref{FIG:FIRST} as a function of $p_\LCm$, with other variables given in the caption\footnote{For simplicity, we rescale to dimensionless variables in all our plots as follows: consider the exponent of (\ref{result}), and then absorb the frequency $k_\LCp$ into the co-ordinates and momenta, so we have $k.x\to x^\LCp$, $k.y\to y^\LCp$, $p_\LCp/k_\LCp\to p_\LCp$, $k_\LCp p_\LCm/m^2 \to p_\LCm$. With this, $x^\LCp$, $y^\LCp$ and $p_\LCpm$ become dimensionless.}. First we demonstrate the convergence of $W_4$ for different values of $\epsilon$: convergence is very rapid for nearly all $p_\LCm$ values as $\epsilon$ decreases. Slower convergence is seen around several sharp edges in $W_4$, which will be explained very shortly. The final result for $W_4$ is represented by the solid line, which we now analyse in more detail. While the free Wigner function is strongly peaked (with a regulator understood) around the free mass-shell condition $p^2 = m^2$, the strong peak in Fig.~\ref{FIG:FIRST} sits at the {\it shifted} mass-shell condition $p^2 = m_*^2$ of the IPW, which is $p_\LCm=\tfrac{13}{4}$ for the given values: like scattering amplitudes, the Wigner function sees particles of free mass $m$ as having mass $m_*$ in an IPW.
Away from the peak, the Wigner function becomes negative: as the shifted mass-shell condition cannot be solved, this naturally  corresponds to quantum (off-shell) effects. For $p_\LCm>\tfrac{13}{4}$, there is in fact no real solution to the deformed mass-shell condition $p^2 = M^2(k.x,k.y)$ for {\it any} $k.x$ or $k.y$. However, for both sufficiently larger and smaller values of $p_\LCm$, the Wigner function becomes positive again. The behaviour of the low $p_\LCm$ values in Fig.~\ref{FIG:FIRST} is reminiscent of the edge and higher harmonic structure in Fig.~\ref{FIG:NLC}, which results from multi-photon effects. Indeed, one may verify that the edges at $p_\LCm=\tfrac{13}{6}$ and $\tfrac{13}{8}$ in Fig.~\ref{FIG:FIRST} are the solutions of the equation (reinserting $k_\LCp$ and $m$ factors)
\be \label{harms}
	(p+nk)^2=m_*^2 \;,
\ee
for $n=1$ and $n=2$. Hence, these edges correspond to resonances in which an {\it off-shell} particle can go on-shell, in the $m_*$ sense, by absorbing photons from the background. The jump at $p_\LCm=\tfrac{13}{2}$, on the other hand, solves (\ref{harms}) for $n=-1$: it corresponds to a particle going on-shell by emitting a photon of laser momentum $k_\mu$.

It is possible to reveal the different processes contributing to the Wigner function by Fourier decomposing the $y^\LCp$ dependence in $M^2$ (as is done to expose the structures in Fig.~\ref{FIG:NLC}, see \cite{Harvey:2009ry}). One finds 
\bea
\label{BesExp}
\nonumber	W_4(x,p) = & \displaystyle\frac{1}{4|p_\LCm|} \displaystyle \sum\limits_n \int\!\ud y^\LCp \exp \bigg[i y^\LCp \bigg( p_\LCp - \frac{p_\LCperp^2 + m^2_*}{4p_\LCm} \bigg)\bigg] \\
	 &\times\exp\bigg[i\frac{n}{2}k_\LCp y^\LCp\bigg]J_n\bigg(\frac{a^2}{k.p}\text{sinc}(\tfrac{k_\LCp y^\LCp}{2})\bigg) \;,
\eea
\begin{figure}[t!]
\centering
\includegraphics[width=0.4\textwidth]{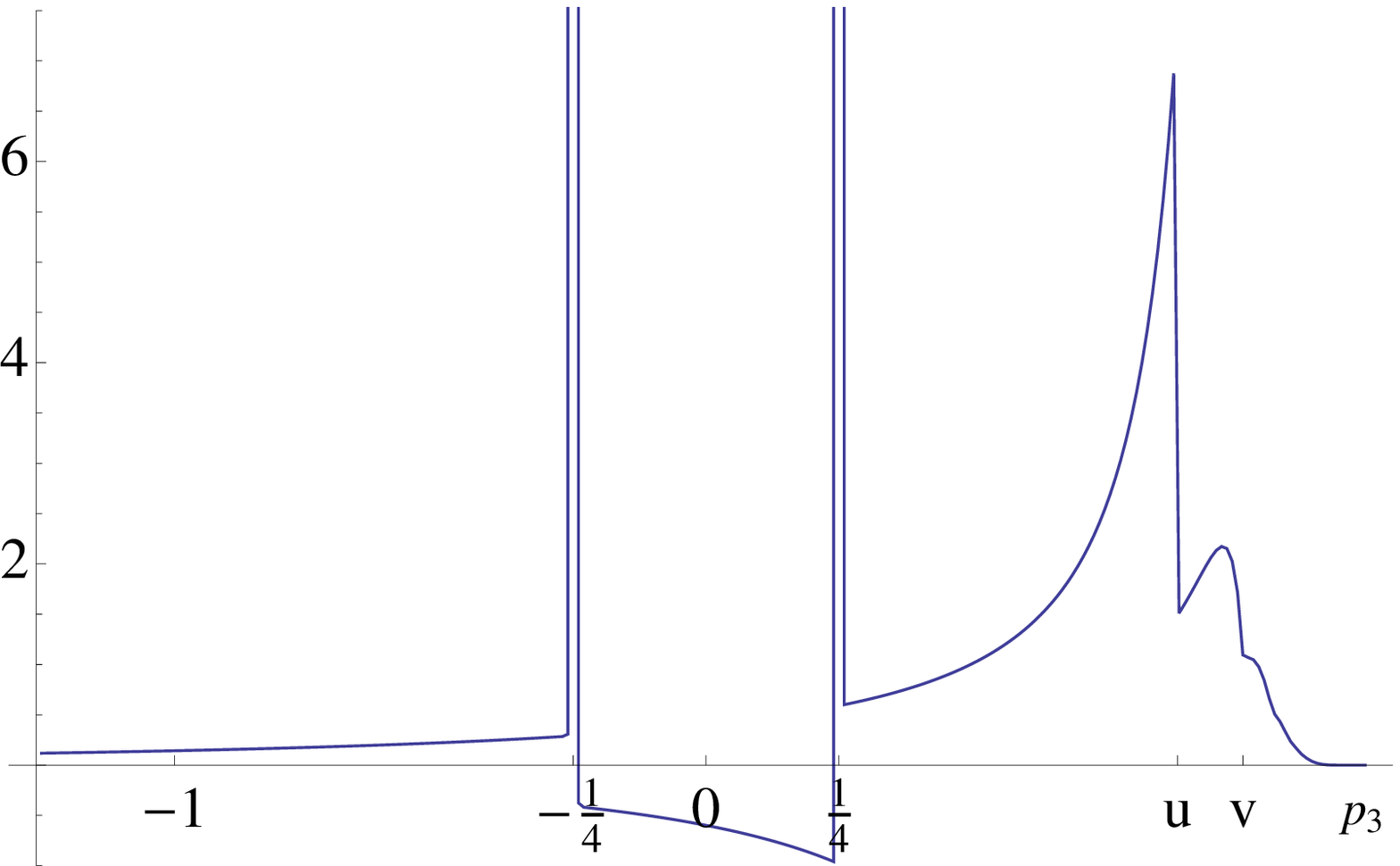}  
\includegraphics[width=0.4\textwidth]{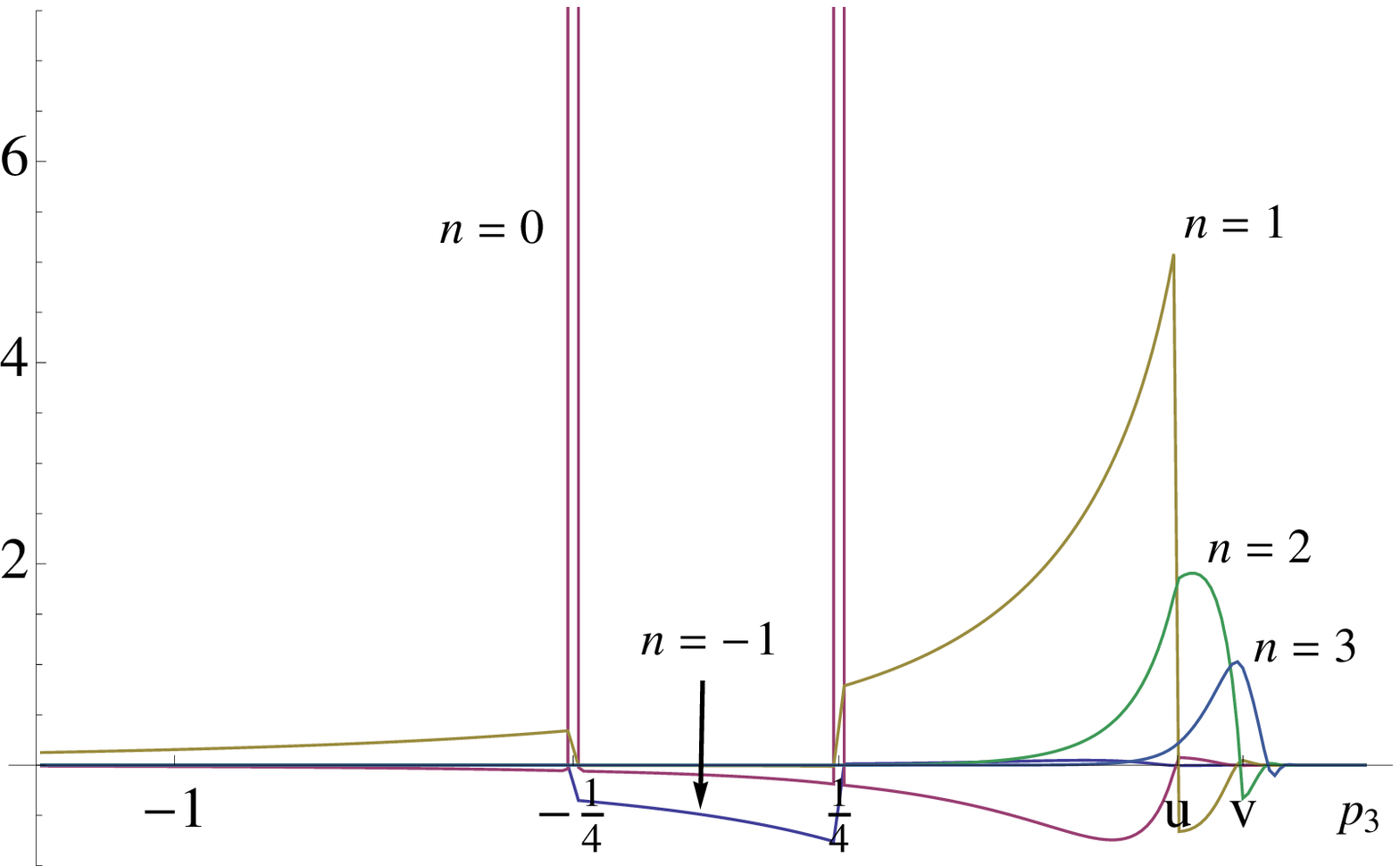}
\caption{\label{FIG:W4} {\it Top}: $W_4$ evaluated at $p_\LCperp=0$, $p_0=5/4$, $a^2=1/2$ as a function of $p_3$, [a.u.]. Quantum effects (negative values) occur when there is no solution to the deformed mass-shell condition.  {\it Bottom}: Contributions from $n=-1$ (emission), $n=0$  and $n=1\ldots 3$ (absorption) photon processes. }
\end{figure}
with $J_n$ Bessel functions of the first kind, and $n$ gives the number of laser photons participating. A particularly neat illustration is shown in Fig.~\ref{FIG:W4}. Here, we use cartesian momentum components, plotting at fixed $p_0$, as a function of longitudinal momentum $p_3$, in order to give an alternative (and perhaps more familiar) view of our results, and a second set of parameters given in the caption. With this choice of parameters, the shifted mass-shell condition $p^2=m_*^2$ corresponds to $p_3 = \pm \tfrac{1}{4}$ in the plots. The Wigner function is negative between these values, and this is precisely where there is no real solution to the deformed mass-shell condition. 

The contributions from $n=-1$ to $n=3$ are shown in the bottom panel of Fig.~\ref{FIG:W4}. One sees that the $n=0$ term is responsible for the strong peaks at the effective mass condition, but is otherwise negative. The $n>0$ terms generate the edge structure: the locations of the first and second edges marked, as `$\mathrm{u}$' and `$\mathrm{v}$' in the plot, correspond to multi-photon conditions (\ref{harms}) for $n=1$ and $2$. Hence the $n$ of the Bessel expansion indeed gives the photon number. The $n=-1$ process contributes largely to the quantum part. We turn now to pulse profiles.
\subsection{Pulse profiles}
We begin with some general properties of $W_4$, which are determined entirely by $M^2$.  As seen in Fig.~\ref{FIG:SHIFT}, a pulse profile brings in a dependence on $k.x$. Here, we focus on the most interesting case, where $k.x$ is chosen to lie in the centre of our pulses.

In a pulse, $M^2$ rises rapidly over the first laser cycle, followed possibly by a brief period of oscillatory behaviour and then a fall back to the rest mass squared. Consequently, we split the integral in (\ref{result}) into two pieces, corresponding to these different behaviours. It is natural (but not necessary) to make this split at the edge of the pulse as seen by the $y^\LCp$ integral, and so we label our two pieces ``in" and ``out". (The change from ``IPW-like" to ``fall-off" behaviour occurs deeper inside the pulse for tighter envelope functions.) The first part of the Wigner function, $W_4^\text{in}$, is clearly always finite. If the pulse envelope varies slowly over most of the pulse duration then $M$ will closely resemble the IPW effective mass; in this case, $W_4^\text{in}$ may be approximately written as (\ref{result}) with $M^2\to M^2_\text{circ}$ and an infra-red cutoff in the $y^\LCp$ integral at the edge of the pulse. $W_4^\text{in}$ then typically exhibits rapid oscillations around the shape of the IPW result.

\begin{figure}[b!!]
\includegraphics[width=0.8\columnwidth]{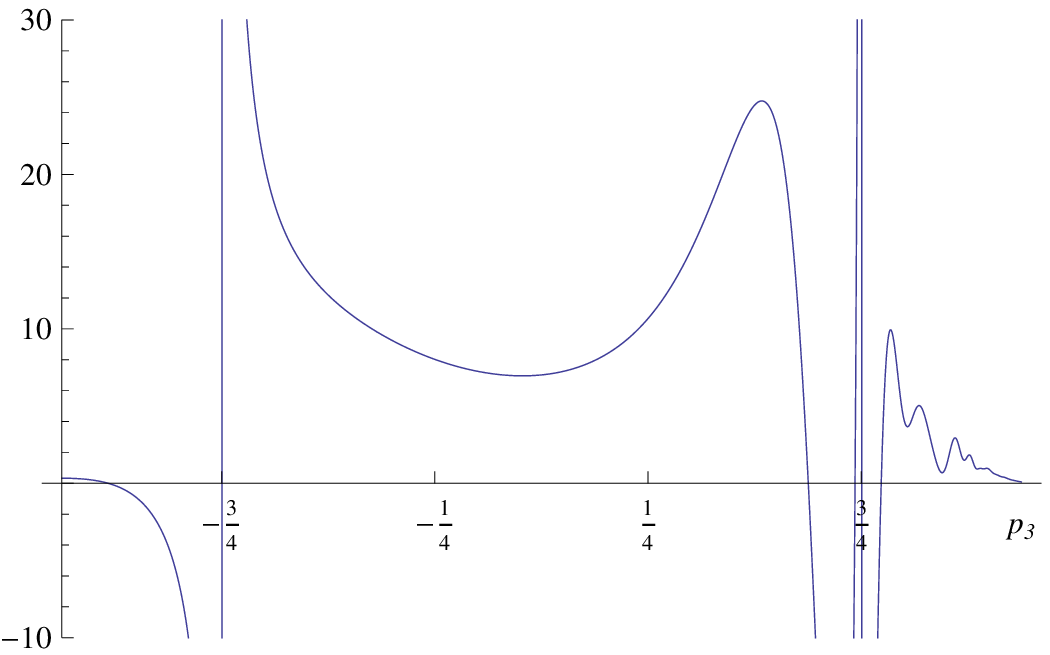}
\includegraphics[width=0.8\columnwidth]{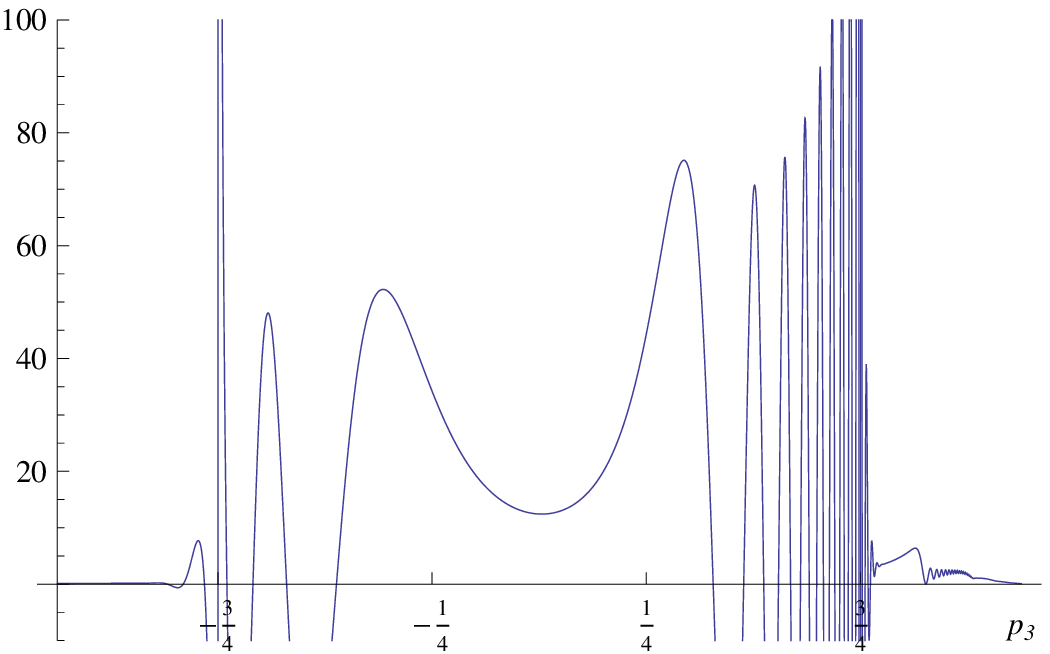}
\includegraphics[width=0.8\columnwidth]{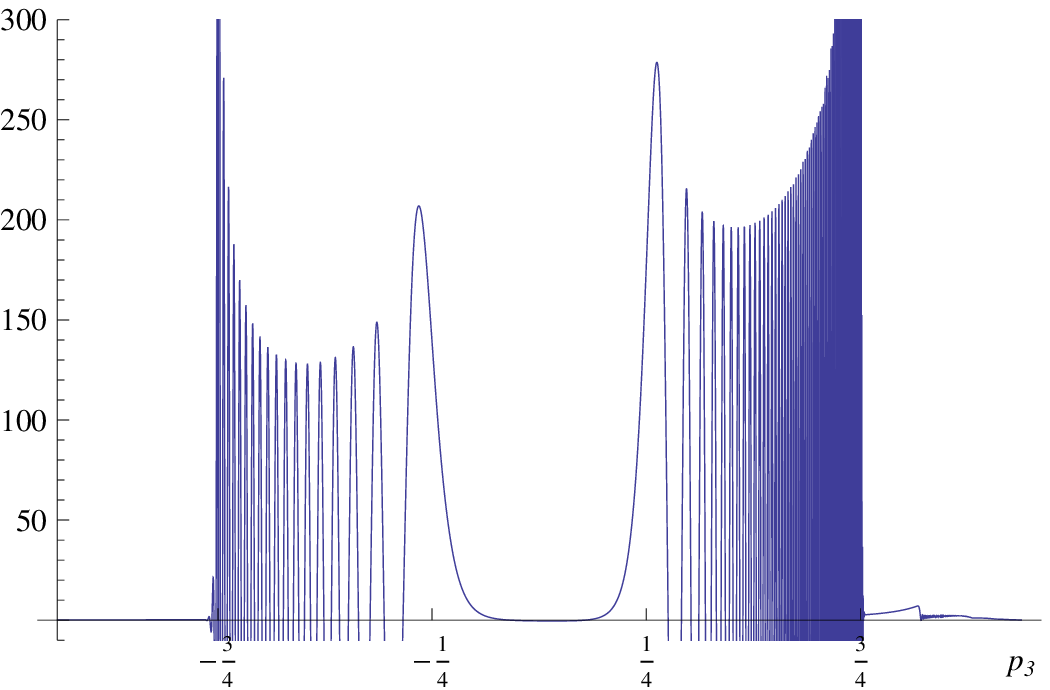}
\caption{\label{FIG:W4Pulse} $W_4$ in the pulse profile (\ref{PulseChoice}), at $p_0=5/4$, $p_\LCperp=0$, $a^2=1/2$, $n=8$, {$64$ and $512$} cycles (descending) with $x^\LCp=n\pi$, i.e.\  in the middle of the pulse. The free mass-shell condition is at $p_3=\pm\tfrac{3}{4}$, the IPW shifted shell is at $p_3=\pm\tfrac{1}{4}$.}
\end{figure}

The remaining part of the integral gives us $W_4^\text{out}$. By construction, $M^2-m^2$ falls off as $1/k.y$ (plus subleading terms) in a pulsed background. This allows us to give an analytic expression for $W_4^\text{out}$:
\be
	W_4^\text{out} = c_1 \delta(p_\LCp - p_\LCp^\text{o.s.})+c_2 \mathcal{P}\frac{1}{p_\LCp - p_\LCp^\text{o.s.}} \;,
\ee
where $p_\LCp^\text{o.s.}$  is shorthand for the on-shell front energy $p_\LCp^\text{o.s.}=(p_\LCperp^2+m^2)/4 p_\LCm$. Special attention should be paid to the appearance of the free mass $m$ in $p_\LCp^\text{o.s.}$. The $c_j$ are finite, pulse-profile dependent functions of momenta, which we do not state explicitly here. Such a contribution is {\it always} found in $W_4$ for pulsed fields as a consequence of the fields eventually turning off, or at least going to zero at large $k.y$. Consequently, we see that infra-red contributions to the Wigner function always yield the free mass-shell condition when we have a pulsed background field. Other mass scales may still play a role in determining the detailed form of the Wigner function (as appears to be the case in scattering calculations \cite{Heinzl:2010vg}).

\begin{figure}[b]
\includegraphics[width=0.85\columnwidth]{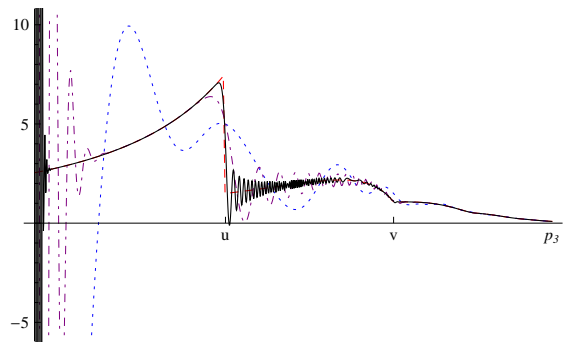}
\caption{\label{FIG:W4Pulse2} The harmonic structure for $n=8$ (dotted), $64$ (dash-dotted) and $512$ (solid), compared with the IPW result (dashed). IPW-like harmonic behaviour appears for a large number of cycles, though the free mass-shell is always present, seen here as the edge at $p_3=\tfrac{3}{4}$ at the very left of the plot.}
\end{figure}

Let us illustrate all this with the pulse  (\ref{PulseChoice}), for which 
\be
	M^2(n\pi, k.y)/m^2 -1 =   a^2\frac{35\pi n}{64}\frac{1}{k.y}
\ee
when $k.y\geq 2\pi n$, i.e. outside the pulse, confirming the falloff behaviour. The Wigner function is plotted in Fig.~\ref{FIG:W4Pulse} for $n=8$, $64$ and $512$ laser cycles. We choose all other parameters as in Fig.~\ref{FIG:W4} to allow easy comparison.  The most striking features are the peaks at the {\it free} mass-shell condition ($p_3=\pm\tfrac{3}{4}$), independent of the pulse length. As the number of cycles increases, rapid oscillations develop around the free mass-shell, while two broad peaks form, and narrow toward the IPW shifted mass-shell at $p_3=\pm\tfrac{1}{4}$. Between these broad peaks, a negative dip forms, which resembles that in Fig.\ \ref{FIG:W4}. However, while this signals the reappearance of IPW behaviour for long pulses, the delta function peaks at the free mass-shell are {\it always} present. 

Hence, the shifted mass itself does not play quite the same role as it does in the IPW: for small numbers of cycles, the shifted mass-shell condition of the IPW is completely absent in the pulsed $W_4$. Even for long pulses, where broad peaks around the shifted mass-shell signify a ``re-dressing" of the electron by the laser field, the Wigner function is always dominated by a delta function imposing the free mass-shell condition.

Toward larger $p_3$ in Fig.\ \ref{FIG:W4Pulse}, one sees a structure similar to the multi-photon effects of the IPW results which exists independent of the pulse length. Because of the scales chosen, this is clearest in the $n=8$ plot, however it is also present for other choices of $n$. In Fig.~\ref{FIG:W4Pulse2} we compare this structure in $n=8$, $64$ and $512$ cycles with the IPW result. One clearly sees that this structure begins to closely resemble the IPW multi-photon structure for a large number of cycles:  note, though, that at the left edge of Fig.\ \ref{FIG:W4Pulse2}, the free mass-shell delta function cuts through the IPW-like structure. This leads us to the conclusion that multi-photon effects remain in the game when we go to pulse profiles, and for very long pulses these learn about the shifted mass $m_*$, as is clear from Fig.\ \ref{FIG:W4Pulse2}, although $m$ continues to play a role.  This is especially interesting in light of the analysis of the SLAC E144 experiment, which confirmed multi-photon effects, but did not unambiguously detect effects stemming directly from the mass shift \cite{McDonald:1999et}.

\begin{figure}[t!!]
\includegraphics[width=0.9\columnwidth]{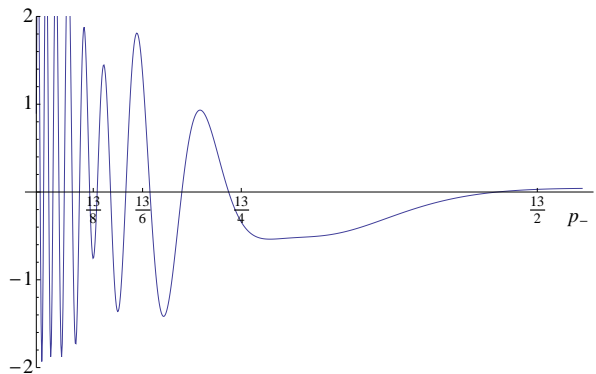}
\caption{\label{FIG:FINAL} $W_4$ in the 8-cycle pulse (\ref{PulseChoice}), as a function of $p_\LCm$.  Other parameters and ranges as in Fig.~\ref{FIG:FIRST} to allow direct comparison. The shifted mass-shell peak is completely absent.}
\end{figure}

We conclude this section with a final comparison between the high intensity IPW result in Fig.\ \ref{FIG:FIRST}, and the Wigner function for a high intensity 8-cycle pulse, shown in Fig.\ \ref{FIG:FINAL}, which further confirms the observations above. We display the same domain and range in both plots for easy comparison. While the large momentum behaviour is reminiscent of the IPW results, including a sign change near the first emission point, the shifted mass-shell at $p_\LCm = \tfrac{13}{4}$ is completely absent. (The free mass-shell is of course present, but lies outside the plotted range.)

\subsection{Equal front time}
We consider now the equal lightfront time Wigner function. To this end, we perform the $p_\LCp$ integration in (\ref{result}), which corresponds to setting $y^\LCp=0$. As the effective mass $M$ deviates from the free mass $m$ due to correlations in front time, one sees that all effective mass effects must vanish when $y^\LCp=0$ and $M=m$. Indeed, one finds that, independent of the pulse profile, $W_3^\LCp = 1/4|p_\LCm|$ just as in the free theory. Hence, calculations of, say, number density at different front times will always agree, and so the triviality of $W_3^\LCp$ seems to reflect the well-known statement that the plane wave cannot produce pairs. It is tempting to conjecture a relation between this result and the `triviality' of the lightfront vacuum \cite{Heinzl:2000ht}.

\subsection{Dirac spinors}
As we have seen, one benefit of the lightfront approach is that the electron-positron Wigner function is particularly simple. It may be the case that one would also like to consider the Wigner function for commutators of the constrained field $\psi_{(\LCm)}$, or of the constrained and dynamical fields. All these interdependent Wigner functions can be neatly represented by the Wigner function defined from the commutator of the full Dirac spinor ${\bra{0}\big[\psi,\bar\psi\big]\ket{0}}$. Writing $\Phi_4$ for the {\it integrand} of the scalar $U_4$, see (\ref{VolkU}), the Dirac field density is
\begin{equation}
\begin{split}
	{\mathcal{U}}_4^\text{Dirac}= \int\!  &\frac{\ud^4q}{{(2\pi)^3}} \Phi_4\bigg[ 1 + \frac{e}{2k.q} \slashed k \slashed A(k.x + \tfrac{k.y}{2}) \bigg]
	\\
	&\times(\slashed q + m) \bigg[ 1 + \frac{e}{2k.q} \slashed A( k.x-\tfrac{k.y}{2}) \slashed k \bigg]  \;,
\end{split}
\end{equation}
where we recognise the spin structure of the fermionic Volkov propagator. The Wigner transform is then performed just as in (\ref{Wig}) and all commutators may be identified by projection onto the relevant $(+)$ and $(-)$ components.

\section{Two approaches to pair production} \label{PPsect}
We briefly outline how the earlier results may be extended to include the effects of vacuum pair production. To do so, we add a second plane wave, i.e.\ a function of $k'.x$ where $k'.k\not = 0$ and vector structure as in (\ref{A}). The collision of two such pulsed plane waves allows us to study vacuum pair production with both electric and magnetic fields, and with finite size effects in {\it two} directions, since pairs are only produced where the fields overlap. 

We consider two experimental setups. The first is the collision of two high intensity lasers, as envisaged for ELI \cite{Bulanov:2010ei}. In this situation, a perturbative field expansion is not permissible. However, if we take the laser pulses to be almost parallel (pair production in parallel pulses is considered in \cite{DiPiazza:2009py}), then $k.k'\approx 0$ because the vector product obeys $\hat{\bf k}.\hat{\bf k}'= 1 +\epsilon\approx 1$, which gives us a small expansion parameter $\epsilon$. Alternatively, for two optical lasers,  we have $k.k'/m^2 \sim 10^{-12}$ independent of their collision angle, which gives us another small parameter with which to construct the Wigner function. Note that both of these approximations are independent of the field strengths: hence the corresponding expansions are nonperturbative in the field intensities, allowing them to be used in the analysis of, for example, multiple ELI-strength lasers. The detailed analysis of these expansions will be a part of our future research.

The second setup is the collision of an intense optical laser with a high frequency, low intensity X-FEL beam. If the amplitude of the latter is sufficiently small, it may be treated in perturbation theory. To illustrate, we begin with the  differential equation defining the Wigner function \cite{Zhuang:1995pd}:
\be
  \label{WigEq}
 \big(\tfrac{i}{2}\slashed D + \slashed \Pi -m \big)\mathcal{W}=0 \; ,
\ee
where the gauge invariant operators $D$ and $\Pi$ are
\be
\begin{split}
  D_\mu &= \frac{\partial}{\partial x^\mu}-e\!\int\limits_{-1/2}^{1/2}\! \ud s\ F_{\mu\nu}(x-is\partial_p)\partial_p^\nu \;,  \\
  \Pi_\mu &=p_\mu-ie\!\int\limits_{-1/2}^{1/2}\!\ud s\ s\,F_{\mu\nu}(x-is\partial_p)\partial_p^\nu \;.
\end{split}
\ee
One now perturbs the field strength, $F_{\mu\nu}\to F_{\mu\nu} + \delta F_{\mu\nu}$, where, for us, $F_{\mu\nu}$ represents the intense (optical) plane wave, and $\delta F_{\mu\nu}$ represents the weak X-FEL field. The above operators are then expanded as $D_\mu\to D_\mu + \delta D_\mu$, etc.  Writing $\mathcal{W}\to \mathcal{W}+\delta \mathcal{W}$, one immediately obtains the equation
\be
  \label{WigEq2}
 \big(\tfrac{i}{2}\slashed D + \slashed \Pi - m\big)\delta \mathcal{W}=-\big(\tfrac{i}{2}\delta\slashed D + \delta\slashed\Pi\big)
 \mathcal{W} \; .
\ee
This equation is solved by inverting the operator on the left hand side to give a Green's function. This may be shown to be a gauge invariant version of the well known Volkov propagator.  Thus, the perturbation $\delta \mathcal{W}$ will describe the emission of electron-positron pairs, dressed by the strong background, from the weak background source. Again, the detailed investigation of this result will be delayed for future work: we mention briefly that the first correction to the equal lightfront time Wigner function, $\delta \mathcal{W}_3^\LCp$, is no longer trivial, which, from our previous discussion, we take as a confirmation of pair production. The Winger function can be seen to depend on both $k'.x$ (which is, say, $x^-$ if we consider a head-on collision) and also the original lightfront direction $k.x$. In particular, the dependence on the strong field remains nonperturbative, as seen in the plane wave $\mathcal{W}_4$. Thus, this approach may also be used to study the role of the effective mass in vacuum pair production.

\section{Conclusions}
We have constructed the Wigner function for both scalar and spinor particles in a background plane wave field exactly. This gives a precise definition of the effective mass which extends the infinite plane wave mass-shift to arbitrary pulse profiles. The Wigner function itself clearly exhibits, in the infinite plane wave case, the shifted mass-shell condition $p^2=m_*^2$, multi-photon absorption/emission, and quantum effects where the effective mass-shell condition has no solution, and the Wigner function becomes negative.

Going to a pulsed profile, we have observed multi-photon structures in both long and short pulses. We have shown that the Wigner function in a pulse always contains, explicitly, a delta function imposing the free mass-shell condition, independent of the pulse length or geometry. This is in contrast to the infinite plane wave case, where the full, nonperturbative Wigner function knows only about the shifted mass-shell.

Finally, we gave various approximations with which to calculate the Wigner function in a more complex field geometry describing colliding laser pulses. This will allow us to study pair creation from the vacuum using the lightfront Wigner approach.

There are many other extensions of the current work which one could consider. For example, transverse size effects are expected to become more important as one goes to higher intensities, since high intensity laser pulses are generated by tight focussing.

Furthermore, recent calculations have shown that pair cascades, through highly focused laser interaction with single electrons, can take place at intensities as low as $10^{24}$ W/cm$^2$ \cite{ELI-report, Elkina:2010up}. Consequently, if large numbers of photons are being taken from the beam, giving rise to significant beam depletion, it will become necessary to go beyond the background field approximation in describing the laser. Although the principles behind the inclusion of effects stemming from beam depletion are reasonably well understood, going beyond the background field approximation still constitutes a major computational challenge.

\acknowledgments
We are grateful to Reinhard Alkofer and Thomas Heinzl for useful discussions. F.~H. is supported by the DOC program of the Austrian Academy of Sciences, and by the FWF doctoral program DK-W1203. A.~I. and M.~M are supported by the European Research Council under Contract No. 204059-QPQV, and the Swedish Research Council under Contract No. 2007-4422.

\end{document}